\documentclass[sigconf]{acmart}

\usepackage{adjustbox}
\usepackage{tcolorbox}
\usepackage{subcaption}
\usepackage{enumitem}
\usepackage{makecell}
\usepackage[nameinlink]{cleveref}
\usepackage{multirow}
\usepackage{threeparttable}
\usepackage{mdframed}
\usepackage{siunitx}
\usepackage{xcolor} 
\usepackage{amsmath}
\usepackage{fancyvrb}
\usepackage{pifont}
\usepackage{xspace}
\usepackage{balance}
\usepackage{framed}
\usepackage{adjustbox}
\usepackage{xurl}
\usepackage{colortbl}

\definecolor{framecolor}{RGB}{0,0,0}
\definecolor{backcolor}{RGB}{245,245,245}
\definecolor{titlecolor}{RGB}{70,70,70}

\mdfdefinestyle{niceframe}{
    linewidth=1pt,
    linecolor=framecolor,
    backgroundcolor=backcolor,
    roundcorner=5pt,
    innertopmargin=\baselineskip,
    frametitlefont=\bfseries\sffamily\color{titlecolor},
    frametitlealignment=\raggedright,
    frametitlebackgroundcolor=backcolor,
    shadow=true,
    shadowsize=3pt,
    shadowcolor=black!20,
}

\newboolean{showcomments}
\setboolean{showcomments}{true}
\ifthenelse{\boolean{showcomments}}
{\newcommand{\mynote}[2]{
\fbox{\bfseries\sffamily\scriptsize#1}
{\small$\blacktriangleright$\textsf{\emph{#2}}$\blacktriangleleft$}}}
{ \newcommand{\mynote}[2]{}}

\newcommand{\wcircle}[1]{\ding{\numexpr171 + #1}}
\newcommand{\bcircle}[1]{\ding{\numexpr181 + #1}}

\newcommand{\dataset}{\textbf{\textsc{{\sc Obscura}}}\xspace}

\newcommand{\finding}[2]{\begin{tcolorbox}[leftrule=0mm,rightrule=0mm,toprule=0mm,bottomrule=0mm,left=0pt,right=0pt,top=0pt,bottom=0pt]
\ding{45} \textbf{[Finding #1]}
#2
\end{tcolorbox}
}

\def\BibTeX{{\rm B\kern-.05em{\sc i\kern-.025em b}\kern-.08em
    T\kern-.1667em\lower.7ex\hbox{E}\kern-.125emX}}

\begin{document}

\title{ The Code Barrier: What LLMs Actually Understand?}

\author{Serge Lionel NIKIEMA}
\orcid{}
\affiliation{%
  \institution{University of Luxembourg}
  \city{Luxembourg}
  \country{Luxembourg}
}
\email{lionel.nikiema@uni.lu}

\author{Jordan Samhi}
\orcid{0000-0001-6052-6184}
\affiliation{%
  \institution{University of Luxembourg}
  \city{Luxembourg}
  \country{Luxembourg}
}
\email{jordan.samhi@uni.lu}

\author{Abdoul Kader Kaboré}
\orcid{0000-0002-3151-9433}
\affiliation{%
  \institution{University of Luxembourg}
  \city{Luxembourg}
  \country{Luxembourg}
}
\email{abdoulkader.kabore@uni.lu}

\author{Jacques Klein}
\orcid{0000-0003-4052-475X}
\affiliation{%
  \institution{University of Luxembourg}
  \city{Luxembourg}
  \country{Luxembourg}
}
\email{jacques.klein@uni.lu}

\author{Tegawendé F. Bissyandé}
\orcid{0000-0001-7270-9869}
\affiliation{%
  \institution{University of Luxembourg}
  \city{Luxembourg}
  \country{Luxembourg}
}
\email{tegawende.bissyande@uni.lu}

\begin{abstract}

Understanding code represents a core ability needed for automating software development tasks. While foundation models like LLMs show impressive results across many software engineering challenges, the extent of their true semantic understanding beyond simple token recognition remains unclear. This research uses code obfuscation as a structured testing framework to evaluate LLMs' semantic understanding capabilities. We methodically apply controlled obfuscation changes to source code and measure comprehension through two complementary tasks: \ding{182} generating accurate descriptions of obfuscated code and \ding{183} performing deobfuscation, a skill with important implications for reverse engineering applications.

Our testing approach includes 13 cutting-edge models, covering both code-specialized (e.g., StarCoder2) and general-purpose (e.g., GPT-4o) architectures, evaluated on a benchmark created from CodeNet and consisting of filtered 250 Java programming problems and their solutions. Findings show a statistically significant performance decline as obfuscation complexity increases, with unexpected resilience shown by general-purpose models compared to their code-focused counterparts. While some models successfully identify obfuscation techniques, their ability to reconstruct the underlying program logic remains constrained, suggesting limitations in their semantic representation mechanisms. This research introduces a new evaluation approach for assessing code comprehension in language models and establishes empirical baselines for advancing research in security-critical code analysis applications such as reverse engineering and adversarial code analysis.

\end{abstract}

\settopmatter{printacmref=false}
\setcopyright{none}
\renewcommand\footnotetextcopyrightpermission[1]{}

\maketitle

\section{Introduction}
The rapid adoption of Large Language Models (LLMs) in software engineering has created a paradox: while these systems can generate syntactically correct code and emulate programming patterns with impressive fluency~\cite{chen2021evaluating, xu2022systematic, tufano2022using}, fundamental questions about their true comprehension capabilities remain unanswered. 
Behind the seemingly successful performance on standard benchmarks lies a critical uncertainty--\textit{are these models genuinely understanding code semantic, or reconstructing patterns encountered during training}?

Code comprehension---the ability to grasp the underlying purpose, logic, and semantics of code---is a fundamental capability for automating software engineering processes~\cite{storey2006tools, maalej2014comprehension}. 
While recent studies have evaluated LLMs on various code-related tasks~\cite{lu2021codexglue, austin2021program, hendrycks2021measuring}, these evaluations typically involve standard, well-formatted code that closely resembles patterns in their training data.
Such evaluations may not effectively distinguish between genuine semantic understanding and sophisticated pattern recognition~\cite{bender2021dangers, marcus2020next}.

This paper introduces a different perspective on assessing LLMs' code comprehension by leveraging code obfuscation as a challenging evaluation paradigm. 
Code obfuscation intentionally transforms code to hinder human comprehension while preserving functionality~\cite{collberg1997taxonomy, lagerkvistdeobfuscation}---creating an adversarial scenario that forces models to demonstrate deeper semantic understanding rather than relying on superficial patterns. 
Unlike previous work that focused on LLMs' performance on standard programming tasks~\cite{chen2021evaluating, xu2022systematic}, our study specifically targets the models' resilience to obscured code structures and their ability to recover semantic meaning despite syntactic distortion.

Recent work explored LLMs' capabilities for security-related issues~\cite{pearce2022asleep}, notably vulnerable code generation, but only few have touched on obfuscated code identification~\cite{fang2023large}. Our work differs significantly by: 
\wcircle{1} systematically applying graduated levels of obfuscation complexity to assess comprehension degradation, 
\wcircle{2} directly comparing code-specialized and general-purpose LLMs across identical tasks, and 
\wcircle{3} evaluating both explanation and deobfuscation capabilities as complementary measures of semantic understanding.

% By answering the study research questions,
With our study, we aim to provide deeper insights into the true nature of LLMs' code comprehension.
This understanding is critical for several reasons:
First, it helps establish realistic expectations about LLMs' abilities, in particular in security-critical applications where adversarial code analysis is essential~\cite{pearce2022asleep}. 
Second, it identifies limitations in current models that inform future research and development~\cite{bommasani2021opportunities}. 
Finally, it establishes a benchmark for measuring progress in LLMs' semantic code understanding over time~\cite{ribeiro2020beyond}.

Our contributions include: 
\begin{enumerate}[noitemsep,topsep=0pt,leftmargin=*]
    \item[\wcircle{1}] identifying obfuscated code analysis as a relevant and yet less-studied task to probe LLMs for code comprehension;
    \item[\wcircle{2}]  an empirical study of 13  state-of-the-art LLMs across diverse obfuscation techniques and complexity levels;
    \item[\wcircle{3}] a comparative analysis revealing unexpected strengths of general models over code-specialized ones on obfuscated code;
    \item[\wcircle{4}] insights into LLMs' practical applications and limitations for deobfuscation tasks in reverse engineering contexts.
\end{enumerate}

\section{Background \& Related Work}
\label{sec:background}
This section provides relevant background on LLMs, current research on code comprehension, and obfuscation techniques.

\subsection{LLMs: Capabilities and Limitations}
\label{subsec:llm-capabilities}

LLMs have demonstrated remarkable capabilities across various domains.
In software engineering, models like GPT-4~\cite{openai2023gpt4}, LLaMA~\cite{touvron2023llama}, and CodeX~\cite{chen2021evaluating} have shown promising results on tasks such as code completion~\cite{svyatkovskiy2020intellicode}, bug fixing~\cite{tufano2022using}, and code translation~\cite{roziere2023code}.

These models generally leverage the transformer architecture~\cite{vaswani2017attention} and self-supervised learning on vast corpora of code, enabling them to capture patterns and relationships in programming languages. Code LLMs, which are code-specialized models such as CodeLLAMA~\cite{rozière2024codellamaopenfoundation} and DeepSeek-Coder~\cite{dai2024deepseekmoeultimateexpertspecialization}, are typically either pre-trained directly on code repositories or fine-tuned from general-purpose LLMs using programming-specific datasets~\cite{xu2022systematic}.

Despite their successes, LLMs exhibit several notable limitations. First, they are susceptible to hallucinations—generating plausible but incorrect or nonsensical outputs~\cite{ji2023survey}. 
Second, they often struggle with complex reasoning tasks that require multi-step logical inference~\cite{wei2022chain}, which is particularly problematic for code understanding. 
Third, their performance degrades significantly on out-of-distribution examples~\cite{yang2022taxonomy}, which suggests reliance on superficial patterns rather than deeper comprehension.

A fundamental challenge in evaluating LLMs for code understanding lies in distinguishing between genuine semantic comprehension and sophisticated pattern matching~\cite{bender2021dangers}. Most existing benchmarks~\cite{lu2021codexglue, hendrycks2021measuring, austin2021program} evaluate models using code samples that closely resemble those in training data, potentially rewarding memorization rather than understanding. As a result, assessing true code comprehension capabilities requires novel evaluation methodologies that can probe beyond surface-level pattern recognition.

\subsection{Studies on LLMs and Code Comprehension}
\label{subsec:code-comprehension-studies}

Code comprehension research with LLMs has evolved across several dimensions. 
Early work focused primarily on syntax-level understanding and code completion tasks~\cite{svyatkovskiy2020intellicode, raychev2015predicting}. 
These studies demonstrated LLMs can predict tokens and complete partial programs but provided limited insight into deeper semantic understanding.

More recent research has explored LLMs' capabilities for higher-level code understanding tasks. 
Studies by Peng et al.~\cite{peng2021could} and Jain et al.~\cite{jain2021contrastive} evaluated models' ability to generate natural language explanations of code functionality, finding that while LLMs can produce plausible explanations for simple programs, their performance deteriorates with increasing program complexity. 
Tufano et al.~\cite{tufano2022learning} investigated LLMs' capabilities for code summarization, demonstrating moderate success but noting significant challenges with larger, more complex codebases.

Several studies have addressed code analysis capabilities of LLMs. 
Prenner and Robbes~\cite{prenner2021automatic} examined how LLMs could detect code smells and suggest refactorings, while Pearce et al.~\cite{pearce2022asleep} evaluated their ability to identify security vulnerabilities. 
Both found that while models showed promising abilities, they missed important issues that human experts would identify. 
Fang et al.~\cite{fang2023large} conducted one of the first studies on LLMs' obfuscation capabilities.

Despite this growing body of research, a comprehensive assessment of LLMs' true code comprehension capabilities remains elusive.
Most studies have focused on standard, well-formatted code rather than challenging models with adversarial examples designed to test deeper comprehension. 
In addition, comparative analyzes between different model architectures and sizes on identical comprehension tasks have been limited.

\subsection{Code Obfuscation Techniques}
\label{subsec:obfuscation-techniques}

Code obfuscation refers to deliberate modifications that make source code difficult for humans to understand while preserving its functionality~\cite{collberg1997taxonomy, lagerkvistdeobfuscation}. 
The motivation of obfuscating software is twofold:
\wcircle{1} to prevent reverse engineering and protect intellectual property; and
\wcircle{2} to hide malicious code from analysts.
While numerous obfuscation techniques exist, our study focuses on three categories that represent different dimensions of code manipulation.

\subsubsection{Lexical Obfuscation: Variable Renaming}
Variable renaming is one of the most common and straightforward obfuscation techniques, targeting the readability of code at the lexical level~\cite{balachandran2013empirical}. This technique involves replacing meaningful identifier names with uninformative alternatives, which eliminates the semantic hints that programmers typically embed in their naming conventions (e.g., replacing variable $accountBalance$ with $a$).

While variable renaming preserves program semantics entirely, it significantly impairs comprehension by eliminating the domain knowledge embedded in identifier names. Studies by Feitelson et al.~\cite{feitelson2020comprehension} have shown that meaningful variable names can improve code comprehension by up to 30\%, making their removal a powerful obfuscation strategy. The idea with this investigation is to assess to what extent LLMs rely on syntactic namings.

\subsubsection{Structural Obfuscation: Dead Code Injection}
Dead code injection alters program structure by adding syntactically valid but semantically irrelevant code segments that are never executed or have no effect on program behavior~\cite{preda2015identifying}. It increases program complexity and obscures control flow and logic by introducing ``noise''. 

Dead code injection significantly increases the cognitive load required for program comprehension by forcing analysts to determine which code segments are relevant to the program's functionality~\cite{meng2013effective}. This technique is particularly effective at confusing both human analysts and automated tools because it requires distinguishing between functional and non-functional code. The idea of this investigation is to assess whether LLMs actually perceive code structure and can disambiguate relevant parts for code behavior. 

\subsubsection{Semantic Obfuscation: Integer and String Encryption}
Integer and string encryption represent semantic obfuscation techniques that transform literal values in the code into encrypted or encoded forms that are decrypted or decoded at runtime~\cite{kovacheva2013analysis}. 
These techniques hide the actual values used in the program, making it more challenging to understand the program's purpose and behavior. 

Unlike variable renaming, which preserves code semantics while changing names, or dead code injection, which adds irrelevant code, encryption transforms the data while maintaining equivalence through runtime decryption~\cite{anckaert2007entropy}. 
This creates a significant challenge for comprehension as it obscures the values in the program, requiring mental execution of the decryption logic to understand program's behavior. 

These three categories of obfuscation--lexical, structural, and semantic--represent complementary approaches.
These constitute a solid foundation for evaluating LLMs' ability to comprehend code despite intentional interferences.
The deliberate opacity introduced by obfuscation creates an ideal test bed for evaluating true code comprehension.
Unlike standard code examples, obfuscated code forces any analyzing entity--human or AI--to look beyond surface syntax and recover semantic  despite intentional interference.

\section{Methodology}
\label{sec:methodology}

This research investigates whether LLMs can comprehend obfuscated code. 
To address this question, we evaluate LLMs on two specific tasks: 
\bcircle{1} explaining obfuscated code; and 
\bcircle{2} generating deobfuscated code.
In this study, we consider the Java programming language as our usage scenario.

\subsection{Research questions}

We aim to answer the following research questions (RQs) to assess LLMs' capabilities to understand the semantics of code:
\begin{itemize}[]
    \item[\textbf{RQ1:}] To what extent can LLMs accurately  describe the semantic purpose of code beyond surface-level syntax? In this RQ, we investigate also how significantly do different levels of code obfuscation impact LLMs' ability to discern a program's underlying functionality.
    \item[\textbf{RQ2:}] How effectively can current LLMs deobfuscate code?
\end{itemize}

\subsection{Benchmarks}
To evaluate LLMs on code semantic understanding, multiple benchmarks are required:
\bcircle{1} first, we need Java source code along with the description of the code;
\bcircle{2} 
second, we need Java source code along with corresponding input--output examples; and 
\bcircle{3} lastly, we need Java source code along with corresponding sets of obfuscated versions at different levels.
We outline our process for constructing our dataset, \dataset, below.

\subsubsection{Initial Dataset}

\begin{table}[htbp]
\centering
\caption{Dataset Comparison}
% \vspace{-5pt}
\begin{adjustbox}{width=.95\columnwidth,center}
\begin{tabular}{lccccc}
\toprule
 & Code Java & Description & Executable & Input/Output & Multiple Solutions \\
\midrule
Codenet\cite{puri2021codenet} & \checkmark & \checkmark & \checkmark & \checkmark & \checkmark\checkmark\checkmark \\
CodeSearchNet\cite{husain2019codesearchnet} & \checkmark & \checkmark & $\times$ & $\times$ & $\times$ \\
Leetcode\cite{zhao2021leetcode} & \checkmark & \checkmark & \checkmark & \checkmark & \checkmark \\
IJA Dataset\cite{fakhoury2019ijadataset} & \checkmark & \checkmark & $\times$ & $\times$ & $\times$ \\
XLCost\cite{guo2021xlcost} & \checkmark & \checkmark & \checkmark & \checkmark & \checkmark \\
CodeXGlue\cite{lu2021codexglue} & \checkmark & \checkmark & \checkmark & \checkmark & \checkmark \\
FunCom\cite{leclair2019funcom} & \checkmark & \checkmark & $\times$ & $\times$ & $\times$ \\
\bottomrule
\end{tabular}
\end{adjustbox}
\label{tab:dataset_comparison}
\end{table}
%%%

We have relied on the CodeNet~\cite{puri2021codenetlargescaleaicode} dataset as our base to build the three datasets we previously described.
Several other datasets could have been suitable, but CodeNet meets all the criteria we previously described and, as shown in Table~\ref{tab:dataset_comparison}, it offers multiple implementations for each problem, paramount for our needs, as we show in this Section.

The CodeNet project is a huge dataset of 14+ million code samples spanning 55 programming languages, created to advance AI code research. 
Sourced from online judge platforms (e.g., AIZU and AtCoder), it contains \num{4053} programming problems.

CodeNet is made of specialized benchmarks such as C++1000, C++1400, Python800, or Java250.
For our research, we have considered the \textbf{Java250} benchmark that contains 250 programming problems with more than \num{75000} possible implementations (i.e., 300 implementations on average per problem).
Note that, for our study, we identified and retained both the most complex implementation of a given problem based on the cyclomatic complexity metric~\cite{mccabe1976complexity}, and the less complex implementation.
Furthermore, we also retained two versions of the source code: 
\wcircle{1} the original source code from CodeNet; and
\wcircle{2} the same piece of code without any comment.

Eventually, our initial dataset, \dataset, is a set of algorithmic problems (e.g., implementing a sorting algorithm), where each problem $P$ consists of $P = (d, \mathcal{I}, s^{+}_{c}, s^{+}, s^{-}_{c})
$, where
\( d \) represents the description of the problem,
\( \mathcal{I} \) is the set of input--output pairs, 
\(s^{+}_{c}\) is the most complex implementation of the given problem, 
\( s^{+} \) is \(s^{+}_{c}\) without the comments in the code, and 
\( s^{-}_{c} \) is the less complex implementation of the given problem.

\subsubsection{Code Description}
\label{sec:code_description}
For a given problem $P \in \dataset$, its description $d$ describes a programming task in the form of a narrative story (e.g., Batman needs to sort a list of Gotham’s most-wanted criminals by threat level, ensuring he takes down the most dangerous ones first.), which might involve specific context, characters, etc.
Moreover, they contain numerical values, constraints, and mathematical formulas. 
As a result, their meaning depends on specific scenarios unrelated to algorithmic logic.
This constitutes an issue to semantically compare LLM-generated descriptions since LLMs would describe the essence of the code without context (e.g., it cannot guess that Batman was actually used to describe the algorithmic problem in a given dataset).

To alleviate this issue, we propose the following solution:
we aim at reformulating any problem description $d$ into another description $d'$ that focused on algorithmic logic.
The goal is to maintain a neutral descriptive structure for fair comparison.
To do so, we use an LLM-based approach to automate the process as well as manual checking to assess the result. 
More specifically, for a given problem $P$, we feed GPT-4o with both $d$ and $s^{-}_{c}$ and ask it to provide free-of-context and descriptive exercise statements $d'$.

This is an example of our prompting strategy to get the free-of-context description of a given problem:

\vspace{0.3cm}
\begin{mdframed}[style=niceframe, linecolor=black, frametitle={Prompt 1},frametitlealignment=\centering]
\vspace{-.4cm}
\textit {"I will provide you with a coding exercise and a corresponding solution. Your task is to analyze both and create a new, beginner-friendly coding exercise. This new exercise should be generic, context-free, and clearly describe the desired functions. Focus on the fundamental programming concepts being practiced and ensure the goal of the exercise is easily understood. It must contain the title of the exercise, the problem statement and the requirement. Do not add example, inputs,outputs. Do not exceed 512 tokens."}
\end{mdframed}
\vspace{0.3cm}

We have \textbf{manually validated all the 250 GPT-4o-generated free-of-context descriptions}.
Among these, 27 did not precisely describe $s^{-}_{c}$.
For these problems, we manually generated the description of the code.
Then, to confirm the similarity of the free-of-context GPT-4o generated description, for each problem $P$, we provide both GPT-4o and Claude Sonnet 3.7 $d$ and $d'$ and ask them to provide a similarity score range from 0 to 1.
Here is an example of the prompt we have used:

\vspace{0.3cm}
\begin{mdframed}[style=niceframe, linecolor=black, frametitle={Prompt 2},frametitlealignment=\centering]
\vspace{-.4cm}
\textit{"You are given two descriptions of code snippets without the code. Provide a similarity score between 0 and 1, where 1 means that the two exercises are similar in terms of goal, constraints, and requirements, and 0 means they solve completely dissimilar."}
\end{mdframed}
\vspace{0.3cm}

For the 250 pairs of $d$ and $d'$, the similarity score ranges from 0.9 to 1.
This means that the reformulated exercise statement kept the meaning and the intention of the original exercise statement.
At the end, we embellish \dataset where each $P$ now consists of $P = (d, d', \mathcal{I}, s^{+}_{c}, s^{+}, s^{-}_{c})$.

\subsubsection{Obfuscation}
To obfuscate the code samples of our dataset into, we employed Obfuscator,  an open-source Java obfuscation tool~\cite{obfuscatorGitHub}.
Obfuscator allows users to select the desired type of obfuscation with different levels of granularity.
For our experiments, we have separately consider the following obfuscation types:
\bcircle{1} variable renaming, which replaces meaningful names with non-descriptive ones which reduces readability while preserving functionality, 
\bcircle{2} dead code injection, which adds non-functional code to obscure comprehension without affecting execution, and 
\bcircle{3} encryption of integer and string literals, which obfuscates literals by replacing them with decryption calls or complex expressions, which increases code complexity while preserving functionality.
For each problem $P \in \dataset$, we considered $s^{+}$

and did the following:

\wcircle{1} we obfuscated $s^{+}$ with variable renaming as $s^{+}_{ren}$;
\wcircle{2} we obfuscated $s^{+}$ with dead code injection as $s^{+}_{dead}$; and 
\wcircle{3} we obfuscated $s^{+}$ with encryption of integer and string literals as $s^{+}_{enc}$.
We, again, embellished \dataset with this information.
Eventually, our dataset \dataset, comprises 250 samples, each described as $P = (d, d', \mathcal{I}, s^{+}_{c}, s^{+}, s^{-}_{c}, s^{+}_{ren}, s^{+}_{dead}, s^{+}_{enc})$.

\subsection{Large Language Models}
\label{sec:methodology:llms}

In our study, we selected various LLMs spanning three main categories. 
The first category consists of pre-trained \emph{generalist} LLMs, such as GPT-4o, which are designed for broad applications across multiple domains.
The second category includes \emph{fine-tuned} LLMs that have been specifically optimized for code-related tasks based on pre-trained foundations, like CodeLlama.
The third category comprises \emph{specialized} LLMs that were purpose-built from the ground up for code-related applications. 
In total, we evaluate 13 distinct LLMs across these three categories.
They are presented in Table~\ref{tab:llms}.

\begin{table}[htbp]
\centering
\caption{Selected Large Language Models}
\label{tab:llms}
\begin{adjustbox}{width=.85\columnwidth,center}
\begin{tabular}{c|l|c|c||c|c}
\toprule
\textbf{ID} & \textbf{LLMs} & \textbf{Weight} & \textbf{Generalist} & \textbf{Fine-tuned} & \textbf{Specialized} \\
\midrule
1 & GPT-4o\cite{openai2024gpt4ocard} & 1760 & \checkmark & & \\
2 & GPT-3.5-turbo\cite{ye2023comprehensivecapabilityanalysisgpt3} & 175 & \checkmark & & \\
3 & 
DeepSeek-R1-Distill-Qwen-7B\cite{deepseekai2025deepseekr1incentivizingreasoningcapability} & 16 & \checkmark & & \\
4 & 
mistralai/Mistral-7B-Instruct-v0.2\cite{jiang2023mistral7b} & 16 & \checkmark & & \\
5 & gemma-2-9b-it\cite{gemma_2024} & 9 & \checkmark & & \\
6 & Llama-3.1-8B\cite{patterson2022carbonfootprintmachinelearning} & 8 & \checkmark & & \\
7 & Qwen2.5-7B-Instruct\cite{yang2024qwen2technicalreport} & 7 & \checkmark & & \\
8 & CodeLlama-7b-hf\cite{rozière2024codellamaopenfoundation} & 7 & & \checkmark & \\
9 & codegemma-7b-it\cite{codegemmateam2024codegemmaopencodemodels} & 7 & & \checkmark & \\
10 & Mamba-Codestral-7B-v0.1 [codestral] & 7 & & \checkmark & \\
11 & DeepSeek-Coder-V2-Lite-Instruct\cite{dai2024deepseekmoeultimateexpertspecialization} & 7 & & \checkmark & \\
12 & starchat2-15b-v0.1\cite{lozhkov2024starcoder2stackv2} & 15 & & & \checkmark \\
13 & starcoder2-15b-instruct-v0.1\cite{wei2024selfcodealignselfalignmentcodegeneration} & 15 & & & \checkmark \\
\bottomrule
\end{tabular}
\end{adjustbox}
\end{table}

\subsection{Evaluation Tasks}
To assess LLMs’ understanding of code semantics, we designed three evaluation phases:
\bcircle{1} generating code descriptions with LLMs and comparing them to our ground truth (Section~\ref{sec:description});
\bcircle{2} generating descriptions for obfuscated code with LLMs and comparing them to our ground truth (Section~\ref{sec:obfuscated_description}); and
\bcircle{3} using LLMs to deobfuscate code and verifying its semantics based on input--output pairs from our ground truth (Section~\ref{sec:deobfuscation}).

\subsubsection{Description generation task}
\label{sec:description}

The code can be in one of two configurations when prompting an LLM for its description:
\wcircle{1} the code contains comments; and
\wcircle{2} the code does not contain any comments.
For this task, we aim at assessing LLMs into generating descriptions in these two configurations and investigate the impact of comments in code samples for LLM code comprehension.

However, not all of the 250 samples in \dataset contain code with comments.
Specifically, not all $s^{+}_{c}$  actually include comments; rather, these samples were not pre-processed to remove comments where present.
As a result, to study the impact of comments, for this task, we only retain any problem $P \in \dataset$ for which $s^{+}_{c}$ actually contains comments, i.e., 139 samples out of 250.

To generate the descriptions of Java code samples, for each of the 139 problems, we prompted each of the 13 LLMs  previously described in Section~\ref{sec:methodology:llms} twice:
\wcircle{1} once with $s^{+}_{c}$, and
\wcircle{2} once with $s^{+}$.

The prompt is as follows:

\vspace{0.3cm}
\begin{mdframed}[style=niceframe, linecolor=black, frametitle={Prompt 3},frametitlealignment=\centering]
\vspace{-.4cm}
"\textit{Your task is to generate a short summary description for a given piece of code. Your description should focus on how the code addresses the problem it was designed to solve.
  Here is how to proceed:
  First, you will be presented with a code snippet: }

  \noindent
    \textbf{\{New Code\}}

    \noindent
  \textit{Next, you will be given an example of code following to his associate problem statement:  }
   \textbf{ \{Java code\}} 
  \textit{And}
   \textbf{ \{Description\}}

 \noindent
\textit{Analyze carefully the \textbf{\{New Code\}}, paying attention to its structure, functions and logic. Consider How each part of the code contributes to solving the given problem.
  Present only your summary description."}
  % within:

  % \noindent
   % \textbf{ \{Response:\}}

\end{mdframed}
\vspace{0.3cm}

In this prompt, \textbf{\{New Code\}} is the piece of code which we want to generate a description by the LLM (i.e., $s^{+}$, and $s^{+}_{c}$), \textbf{\{Java Code\}} is another piece of code serving as an example for the LLM along with its description \textbf{\{Description\}} to guide the LLM.

Next, for each problem $P$, we compare each LLM-generated description to $d'$ using the same prompt as in Section~\ref{sec:code_description}.
That is to say, we ask both GPT-4o and Claude Sonnet 3.7 for a similarity score between 0 and 1.
If any of the LLM yields a score below 0.5, we consider the LLM-generated description as not accurately describing $s^{+}$ or $s^{+}_{c}$.

%\vspace{-.1cm}

\begin{figure}[htbp]
    \centering
    \begin{adjustbox}{width=.75\columnwidth,center}
    \includegraphics[width=0.5\textwidth]{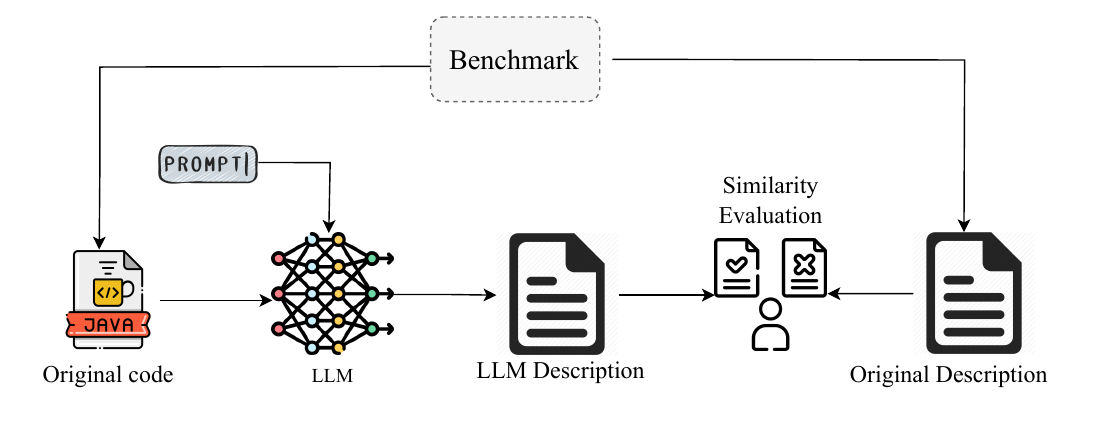} 
    \end{adjustbox}
    \vspace{-.5cm}
    \caption{Description generation process and comparison against ground-truth}
    \label{fig:Description}
\end{figure}

For this task, we used the following prompt:
\vspace{0.1cm}
\begin{mdframed}[style=niceframe, linecolor=black, frametitle={Prompt 4},frametitlealignment=\centering]
\vspace{-.4cm}
\textit{"You are given two descriptions of code snippets without the code. Provide a similarity score between 0 and 1, where 1 means the two description is about code that solve the same problem, and 0 means they solve completely different problem. Respond with only a numerical value between 0 and 1."}
\end{mdframed}
\vspace{0.3cm}
The overview of this process is shown in Figure~\ref{fig:Description}.
We present the results in Section~\ref{sec:results:baseline}.

\subsubsection{Obfuscated code description generation task}
\label{sec:obfuscated_description}
To generate descriptions of obfuscated Java samples, for each problem $P \in \dataset$, we prompt LLMs to first deobfuscate $s^{+}_{ren}, s^{+}_{dead}, s^{+}_{enc}$ and then GPT-4o to generate a description for each of the deobfuscated samples.
The prompt is as follows:

\vspace{0.3cm}
\begin{mdframed}[style=niceframe, linecolor=black, frametitle={Prompt 5},frametitlealignment=\centering]
\vspace{-.4cm}
\textit{You are an expert in reverse engineering. You will be given an obfuscate code. Your task is to generate a readable and well formatted version of this code.}

\noindent
\textit{First, analyze thoroughly the obfuscated code to understand its structure and functionality.}

\noindent
\textbf{\{Obfuscated Code\}}

\noindent
\textit{Generate a deobfuscated version using meaningful variable and method names, apply proper indentation, follow Java naming conventions, simplify complex expressions, remove redundant code, maintain the original functionality.}

\noindent
\textit{Return only the readable code.}
% inside:

% \textbf{\{Deobfuscate code\}}}

\end{mdframed}
\vspace{0.3cm}

Next, for each LLM-generated description of deobfuscated code, we use the same prompt as in Section~\ref{sec:description}.
That is to say, we ask both Claude Sonnet 3.7 and Deepseek-reason for a similarity score between 0 and 1.
If any of the LLM yields a score below 0.5, we consider the LLM-generated description as not accurately describing $s^{+}$.
We present the results in Section~\ref{sec:results:obfuscation}.

\subsubsection{Deobfuscation Task}
\label{sec:deobfuscation}
To assess LLMs' ability to derive meaningful and readable versions of obfuscated code, we reuse the deobfuscated Java samples previously described in Section~\ref{sec:obfuscated_description}.

\begin{figure}[htbp]
    \centering
    \begin{adjustbox}{width=.9\columnwidth,center}
    \includegraphics[width=0.5\textwidth]{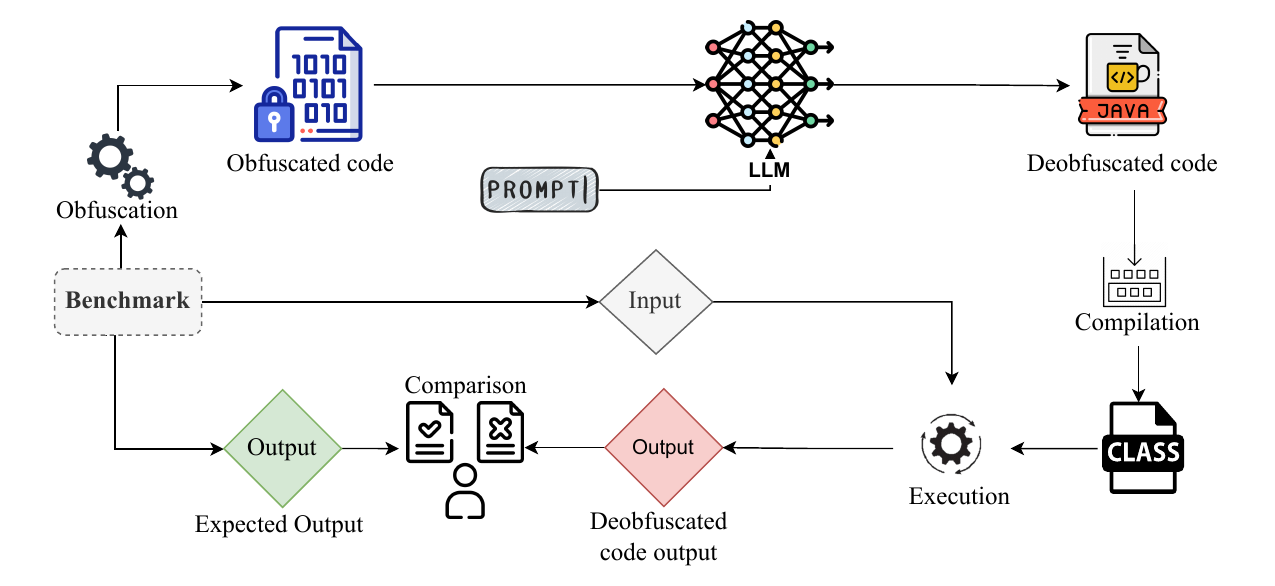} 
    \end{adjustbox}
    % \vspace{-20pt}
    \caption{Deobfuscation process}
    \label{fig:Debobfuscation process}
    \vspace{-0.5cm}
\end{figure}

Then, for each problem $P$, we compared each LLM-generate deobfuscated piece of code and perform the following steps:
\bcircle{1} we attempt to compile the deobfuscated code;
\bcircle{2} if compilation succeeds, we attempt to execute it;
\bcircle{3} if the program does not crash during execution, we use our set of input--output pairs $\mathcal{I}$ to run the program with the inputs and collect the outputs; and
\bcircle{4} we compare the generated output (after execution) with the expected outputs from our ground-truth $\mathcal{I}$.
The overview of this process is shown in Figure~\ref{fig:Debobfuscation process}.
We present our results in Section~\ref{sec:results:obfuscation}.
% \vspace{-5pt}

\subsection{Evaluation Metrics}

During this study, we employed several metrics to evaluate the effectiveness of LLMs in generating descriptions and their ability to provide deobfuscated code:
\begin{itemize}[leftmargin=*]
    \item Cyclomatic Complexity: We tracked the evolution of code complexity before and after obfuscation, providing a quantitative measure of structural changes.
    \item LLM-Based Semantic Similarity: To overcome limitations of traditional methods like embedding with vector similarity computation, we implemented an LLM-based measurement approach. We use three 
    
    models to evaluate the semantic similarity between text (i.e., GPT-4o for description generation of deobfuscated code, and Deepseek-reason and Claude for evaluating descriptions semantic accuracy).

    \item Statistical Analysis: We employed the Mann-Whitney-Wilcoxon (MWW) \cite{mann1947test} test to measure the statistical differences between samples where LLMs failed to generate adequate descriptions versus the original dataset, analyzing differences in complexity, lines of code, and token count.
\end{itemize}{}

As a supplementary validation measure, we prompted these same two models to provide binary (yes/no) classifications regarding whether the generated descriptions were programming-related. This additional step helped verify the fundamental relevance of the generated content to the programming domain.

\section{Experimental Results}
\label{sec:results}

This section presents the experimental results of our study.

\subsection{Baseline Analysis: Comment and Language Effects on LLM Code Description Generation}
\label{sec:results:baseline}

In this section, we present the results of our evaluation of LLM-generated descriptions on non-obfuscated code.
Section~\ref{sec:results:comments} investigates the impact of comments, and Section~\ref{sec:results:languages} investigates the impact of the language used in comments.

\subsubsection{Impact of Comments in Code for LLM Code Comprehension}
\label{sec:results:comments}
\noindent
\textbf{Goal:} We investigate how comment tokens influence LLMs when generating code descriptions. Despite comments being relatively modest in size in our dataset, they may significantly impact the LLMs' outputs compared to relying solely on the lexical tokens of the code itself. We assess the relative weight LLMs assign to natural language explanations versus code structure and syntax.

\noindent
\textbf{Experiments:} 
We identified 101 problems in our dataset where the implementation ($s^{+}_{c}$) contains comments in the English language. For these problems, we prompted each LLM to generate descriptions for both the commented version ($s^{+}_{c}$) and the version with comments removed ($s^{+}$). 
We then evaluated the accuracy of these descriptions using the process described in Section~\ref{sec:code_description}.

\begin{figure}[!h]
    \centering
    \begin{adjustbox}{width=1\columnwidth,center}
    \includegraphics[scale=1]{
    %Figures/With_Whithout_Comment.pdf
    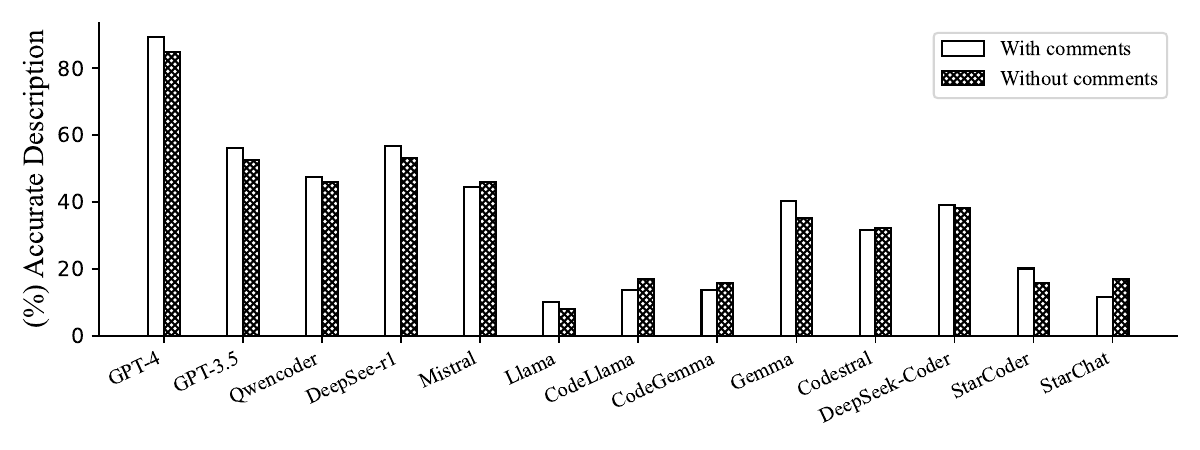
    } 
    \end{adjustbox}
    \vspace{-25pt}
    \caption{LLMs effectiveness in code description (with or without comments)}
    \label{fig:WithvsWithoutcomments}
\end{figure}

\noindent
\textbf{Results:} 
Figure~\ref{fig:WithvsWithoutcomments} presents the percentage of accurate descriptions generated by each LLM under both conditions. We observe:
\begin{enumerate}[leftmargin=*]
\item For 8 out of 13 LLMs tested, the performance in generating accurate descriptions is higher when considering code with comments. The average improvement was 3.7\%, with Deepseek-R1 showing the largest gain (11.5\%).
\item GPT-4o drastically outperforms all other LLMs tested in our study, achieving around 87\% accuracy with comments and around 84\% without comments. The differences in performance distributions are however not statistically significant (MWW).

\item There is a striking performance disparity across models, with some code-specialized models surprisingly underperforming compared to general-purpose models. Notably, several models (Mistral, CodeLlama, CodeStral, StarChat) exhibit the counterintuitive pattern of performing better with code without comments, showing although only small improvements of 2.9, 5.0, 2.9, and 2.2 percentage points, respectively.
\end{enumerate}

We speculate that this performance variation might still reflect fundamental differences in model training and/or architecture. Models showing improvement with comments 
 likely leverage the natural language explanations as additional context, while those performing better without comments may have developed stronger representations of code syntax and structure during training, potentially being distracted by comments that don't align with their learned patterns. However, given the complexity of these models and the limited scope of our experiment, multiple alternative explanations may exist, including differences in tokenization approaches, varying sensitivities to comment placement, etc. 
% \vspace{-10pt}

\vspace{-.6cm}

\finding{1}{
Our empirical results show general-purpose LLMs outperforming code-specialized models in description accuracy.  This challenges the common assumption in code-specialized model development that domain-specific training necessarily enhances performance on code tasks. The data suggests that foundational model quality and/or pre-training diversity may be more significant factors than domain specialization for code comprehension tasks.
On the studied benchmarks, the effect of comments was not shown statistically significant, potentially due to the fact that these were not Javadoc-style documentation, but rather relatively-short comments (cf. Figure~\ref{fig:Ratio}).
}

\subsubsection{Impact of Comment Language on LLM Code Understanding}
\label{sec:results:languages}

\noindent
\textbf{Goal:} We investigate how the natural language (e.g., English) in which code comments are provided affects LLMs' code comprehension abilities. While programming languages are standardized, comments can be written in any natural language.

\noindent
\textbf{Experiments:}
We isolated from our dataset all problems whose reference solutions are associated with Japanese comments. Then we used GPT-4o to translate the Japanese comments to English. We then compared the accuracy of descriptions generated from code with the original Japanese comments versus code with the translated English comments.

Additionally, we analyzed the comment density (ratio of comment lines to total code lines) in the original sets of problems with English and Japanese\footnote{We translate the Japanese comments into English to ensure unbiased comparison given lexical differences.} comments to better understand potential differences in documentation practices.

\begin{figure}[htbp]
    \centering
    \begin{adjustbox}{width=1\columnwidth,center}
    \includegraphics[scale=1]{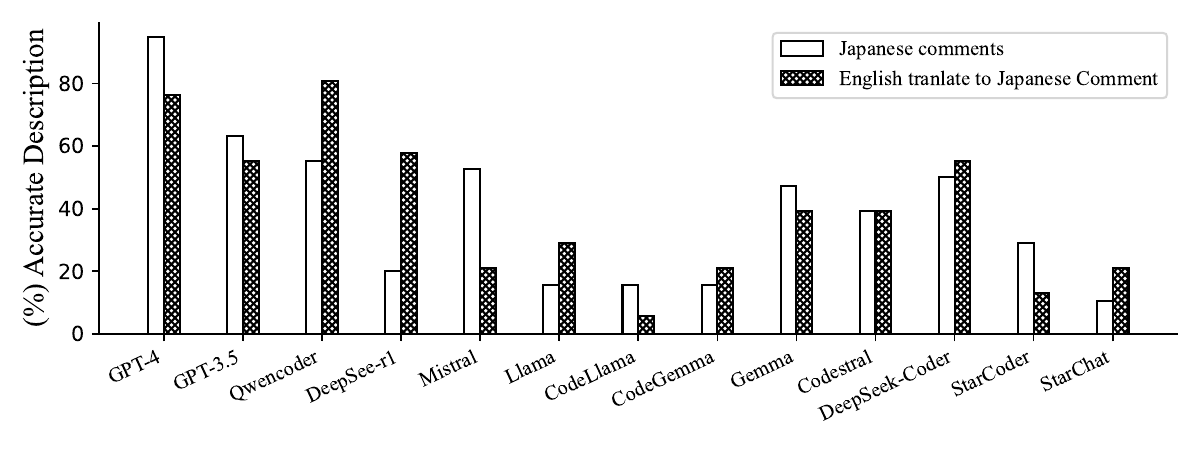} 
    \end{adjustbox}
    % \vspace{-25pt}
    \caption{Effectiveness of LLMs to generate descriptions on problems with comments (Japanese vs English translations).}
    \label{fig:Japanesetraduct}
\end{figure}

\noindent
\textbf{Results:}
Figure~\ref{fig:Japanesetraduct} shows the results of our experiment with Japanese comments. Six LLM models show improved performance when processing code with translated comments compared to the original Japanese comments. Among these, Deepseek-R1 demonstrated the most significant improvement, with a 37.5\% increase in effectiveness when working with translated comments compared to the original Japanese comments.
% \vspace{-10pt}

\begin{figure}[!h]
    \centering
    \begin{adjustbox}{width=.95\columnwidth,center}
    \includegraphics[scale=1]{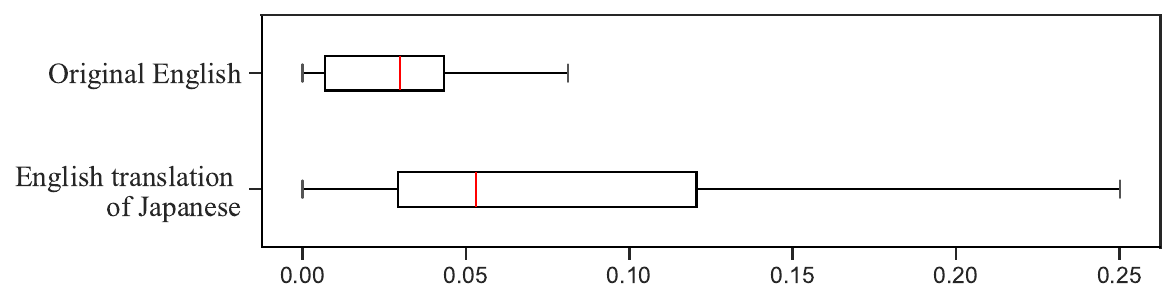} 
    \end{adjustbox}
% \vspace{-20pt}
    \caption{Comment to code Ratio-English vs Japanese (translated in English)}
    \label{fig:Ratio}
\end{figure}

Figure~\ref{fig:Ratio} shows the comment density distribution in both sets of problems. Code samples with Japanese comments (after translation) had significantly higher comment ratios—up to 25\% of total lines—compared to samples with native English comments, which peaked at approximately 7-8\%. This difference in documentation density also likely contributed to the substantial performance improvements observed after translation, as more comprehensive comments provided richer context for the models to leverage.

\finding{2}{
Our results reveal nuanced patterns in how comment language affects LLMs: (1) Models show varied responses to comment language—some perform better with original Japanese comments (GPT-4, GPT-3.5, Mistral, QwenCoder), while others improve with English translations (DeepSeek-r1, Llama, CodeGemma, StarChat); (2) Japanese comments contain significantly higher comment density (median ~10\% vs ~5\% in English), suggesting potential cultural differences in documentation practices. These findings highlight both the diverse multilingual capabilities of different LLMs and the importance of comment comprehensiveness in code understanding tasks.
 }

\subsection{Effect of Obfuscation on LLM Code Description Generation}
\label{sec:results:obfuscation}
In this section, we present our evaluation of how different obfuscation techniques affect LLMs' ability to generate accurate code descriptions. Section~\ref{sec:results:obfuscation:performance} assesses performance across obfuscation types, Section~\ref{sec:results:obfuscation:fail} investigates characteristics of consistently misunderstood code, and Section~\ref{sec:results:obfuscation:code_related} examines LLM failures' nature.

\subsubsection{LLM-Generated Description Performance of Obfuscated Code}
\label{sec:results:obfuscation:performance}

\noindent
\textbf{Goal:} We investigate how different obfuscation types affect LLMs' ability to generate accurate code descriptions. 
This comparison is paramount because obfuscation targets different aspects of code (i.e., lexical, structural, semantic), allowing us to isolate which code properties LLMs rely on most when processing programs.

\noindent
\textbf{Experiment:}
For each problem in \dataset, we prompted the 13 LLMs to generate descriptions for each obfuscated variant: variable renaming ($s^{+}{ren}$), dead-code-injected ($s^{+}_{dead}$), and enc\-rypted-literals ($s^{+}_{enc}$). 
We evaluated description accuracy using the same process as for non-obfuscated code. 
By comparing performance between obfuscated and non-obfuscated variants, we can quantify how each obfuscation type impacts LLMs' description capabilities.

\noindent
\textbf{Results:}
Figure~\ref{fig:obfuscation-impact} presents the accuracy of descriptions generated for each obfuscation technique compared to the non-obfuscated baseline. We observe consistent performance degradation across most models, with technique-specific patterns:

\begin{figure*}[!t]
\centering
\includegraphics[width=1\textwidth]{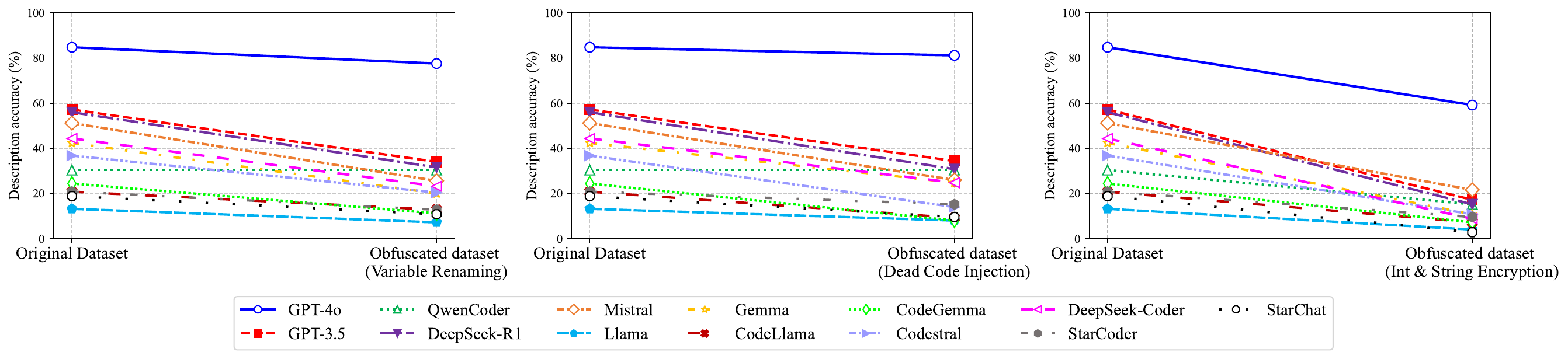}
\caption{Impact of obfuscation techniques on description accuracy compared to non-obfuscated code - \normalfont comments are removed from all samples in the original dataset and before any obfuscation technique is applied.}
\label{fig:obfuscation-impact}
\end{figure*}

\noindent {\textit{Variable Renaming} caused significant performance degradation across almost all models. The average accuracy dropped by 18.6 percentage points, with some models experiencing declines of over 30 percentage points. This pronounced sensitivity to identifier names suggests LLMs heavily rely on lexical semantics embedded in variable names rather than structural code understanding. GPT-4o showed the most resilience, with only a 7.3\% decline.

\noindent \textit{Dead Code Injection} had, surprisingly, the least impact on description accuracy, with an average decline of only 6.2\%. Several models (QwenCoder, Mistral, and GPT-4o) showed minimal performance changes, suggesting they can effectively identify and filter out non-functional code segments when generating descriptions.

\noindent \textit{Literal Encryption} caused substantial performance degradation (average decline of 21.4\%), particularly affecting ge\-neral-purpose models. Even GPT-4o experienced a significant drop of 19.9\%, though it remained the top performer with 58.8\% accuracy on encrypted code. This suggests that obfuscating literal values, associated with the introduction of complex code for run-time decryption, severely impairs LLMs' ability to infer program semantics.

QwenCoder showed unusual resilience to variable renaming and dead code injection compared to other models, maintaining nearly identical performance across these conditions. 
This outlier behavior warrants 
% further
investigation into its training data and/or architecture.

\finding{3}{
LLMs exhibit critical dependencies on semantic elements: meaningful variable names and unobfuscated literals.
Description accuracy drops significantly when these human-readable components are altered, while structural obfuscation through dead code has less impact. 
This suggests current models process code primarily through surface-level linguistic patterns rather than deeper algorithmic structures independent of representation.
}

\subsubsection{Why LLMs Fail with the Code Description Generation Task}
\label{sec:results:obfuscation:fail}

\noindent
\textbf{Goal:} We investigate the reason why LLMs struggle to generate code descriptions for obfuscated code and non-obfuscated.

\noindent
\textbf{Experiment:}
We identified 13 samples where all LLMs failed to generate accurate descriptions across all conditions (non-obfuscated and all three obfuscation variants). For these consistently misunderstood samples, we calculated cyclomatic complexity and token count, then compared their distributions against the entire dataset using the Mann-Whitney U test to determine statistical significance.

\noindent
\textbf{Results:}
Figure~\ref{fig:Complex:tokens} shows the distribution of cyclomatic complexity and token count for both the failed samples and the original dataset. The analysis reveals a striking pattern:

Failed samples exhibit higher average and median cyclomatic complexity values (38.7 vs. 32.5 for average and 31 vs. 37 for median) while containing substantially fewer tokens (405.5 vs. 664.0 on average and 609 vs. 364 in terms of median). The token distribution in failed samples also shows a narrower range (standard deviation 171.4 vs. 336.5). The Mann-Whitney U test confirmed statistical significance for token count differences (p-value = 0.007), though the complexity difference was not statistically significant.
Nevertheless, the perceived inverse relationship between complexity and code size suggests that LLMs struggle most with code that packs complex logic into compact implementations. These dense code structures, which do "more with less", present a particular challenge for current model architectures.

\begin{figure}[htbp]
    \centering
    \begin{adjustbox}{width=.9\columnwidth,height=0.25\columnwidth,center}
\includegraphics[width=0.5\textwidth]{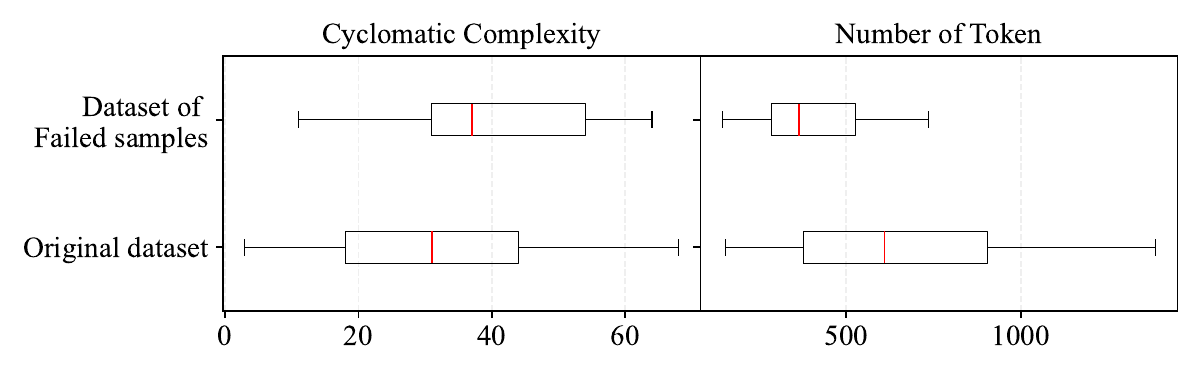}
    \end{adjustbox}
    \caption{Distribution of cyclomatic complexity and token counts for samples where all LLMs failed to generate an accurate description, compared to the entire dataset.}
    \label{fig:Complex:tokens}
\end{figure}

\finding{4}{
 Samples with high information density—implementing complex logic with fewer tokens—present a unique challenge for current LLM architectures. This finding suggests a blind spot in model capabilities when processing code that combines brevity with logical complexity.
}

\subsubsection{Do LLMs Hallucinate or Misinterpret Code?}\hfill
\label{sec:results:obfuscation:code_related}

\noindent
\textbf{Goal:}
We investigate to what extent LLMs generate description content that is entirely unrelated to coding, such as a poem, which would indicate hallucination.

\noindent
\textbf{Experiment:}
To achieve our goal, we relied on GPT-4o and prompted it for all descriptions generated by all LLMs on all obfuscated variants of the 250 Java samples. The prompt is as follows:

\vspace{0.3cm}
\begin{mdframed}[style=niceframe, linecolor=black, frametitle={Prompt 6},frametitlealignment=\centering]
\vspace{-.4cm}
\textit{"Determine if the given code description is programming-related or not. Answer with only `yes' or `no'."}
\end{mdframed}
\vspace{0.3cm}

The rationale is as follows:  
If the generated description is related to code, it suggests that it simply does not accurately describe the sample, i.e., it is a misinterpretation rather than a complete failure.  
However, if the description is entirely unrelated to code, it indicates that the model has hallucinated while generating the description.

\begin{figure}[htbp]
    \centering
    \begin{adjustbox}{width=.9\columnwidth,center}
    \includegraphics[scale=1]{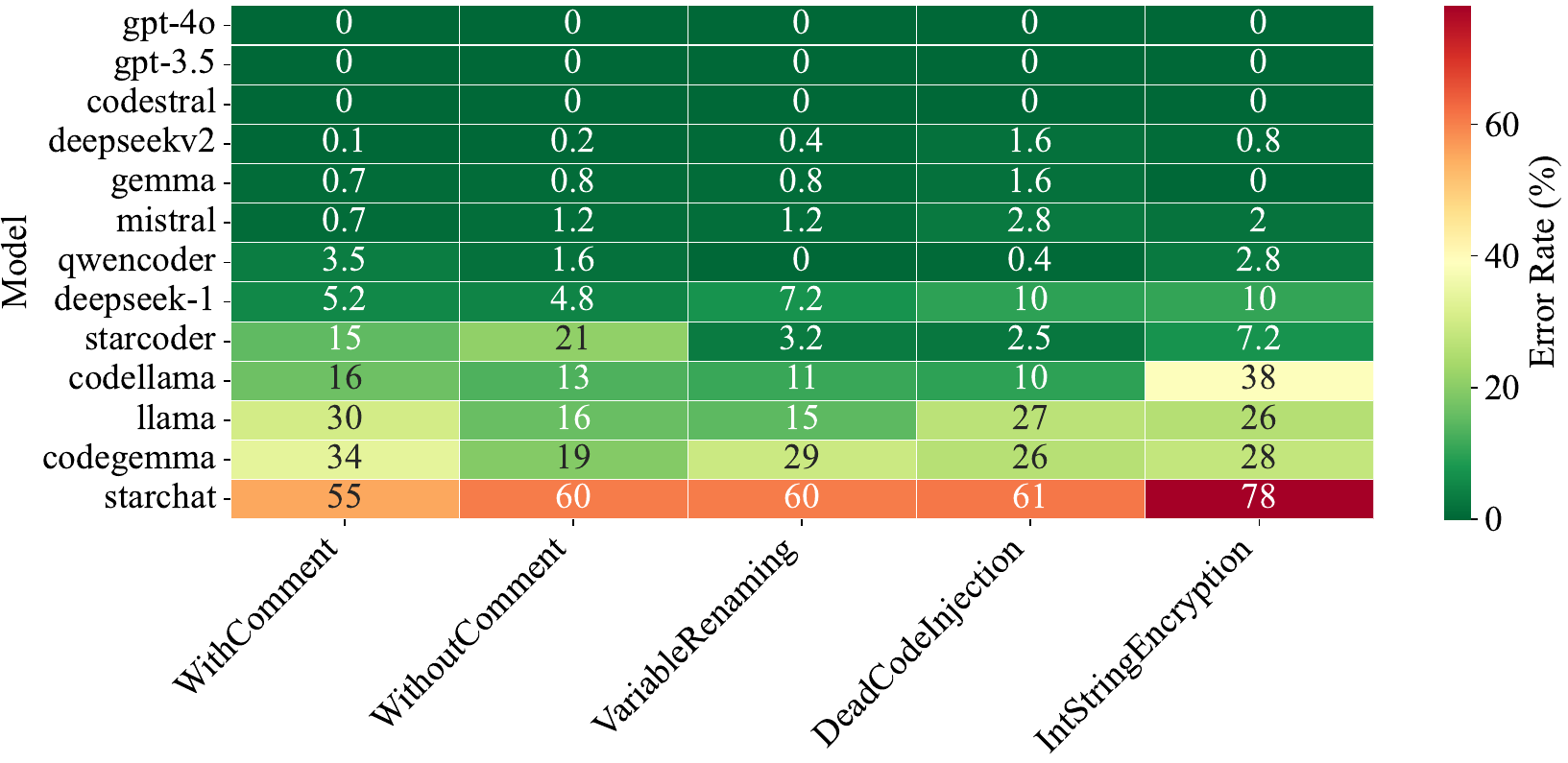} 
    \end{adjustbox}
    \vspace{-20pt}
    \caption{LLMs Error Rate of Generating Code-related Description}
    \label{fig:ErrorRate}
\end{figure}

% \vspace{-1cm}

\noindent
\textbf{Results:}
Figure~\ref{fig:ErrorRate} presents our results.
It includes the error rate, of the studied LLMs, in generating \emph{code-related} descriptions across all code sample configurations.
Several models, despite being unable to provide accurate descriptions, consistently generate content that remains within the programming domain, i.e., code descriptions are related to code. 
This suggests that these models maintain contextual awareness even during failure modes.
Conversely, other models demonstrate a tendency to produce descriptions that completely diverge from the programming context when they fail, i.e., the descriptions generated are unrelated to code. 
An unexpected finding is that one of the models exhibiting the highest rate of contextually unrelated descriptions is specifically designed and fine-tuned for code-related tasks.

\finding{5}{
Our analysis reveals significant variation in how LLMs fail when generating descriptions. Top-performing models (GPT-4o, GPT-3.5, and CodeStral) never produce non-code-related content, while other models show increasing rates of domain-irrelevant outputs. Surprisingly, StarChat—specifically designed for code tasks—exhibits the highest error rates (55-78\%) in generating code-related descriptions across all configurations. This suggests fundamental differences in how models maintain contextual awareness under challenging conditions, with some preserving domain relevance despite inaccuracies while others experience catastrophic failures producing completely unrelated content.
}

\subsection{Deobfuscation Capabilities}
\subsubsection{Can LLMs effectively deobfuscate code?}\hfill
\label{deobfs}

\noindent
\textbf{Goal:} 
We evaluate how effectively LLMs can reconstruct clean, functionally correct code from obfuscated inputs. This capability represents a stronger test of code comprehension than description generation, as it requires not just understanding the code's purpose but also reconstructing its implementation details.

\noindent
\textbf{Experiment:} 
We feed obfuscated code samples to LLMs with a specialized prompt (Prompt 5) requesting deobfuscated versions, then we validate the generated solutions through an automated pipeline that compiles the code, tests it with dataset inputs, and compares outputs against expected results. This approach quantitatively measures how effectively LLMs can extract meaningful, functionally equivalent implementations from deliberately complex code, providing insights into their semantic understanding and code refactoring capabilities.

\begin{figure}[htbp]
    \centering
    \includegraphics[width=.9\linewidth]{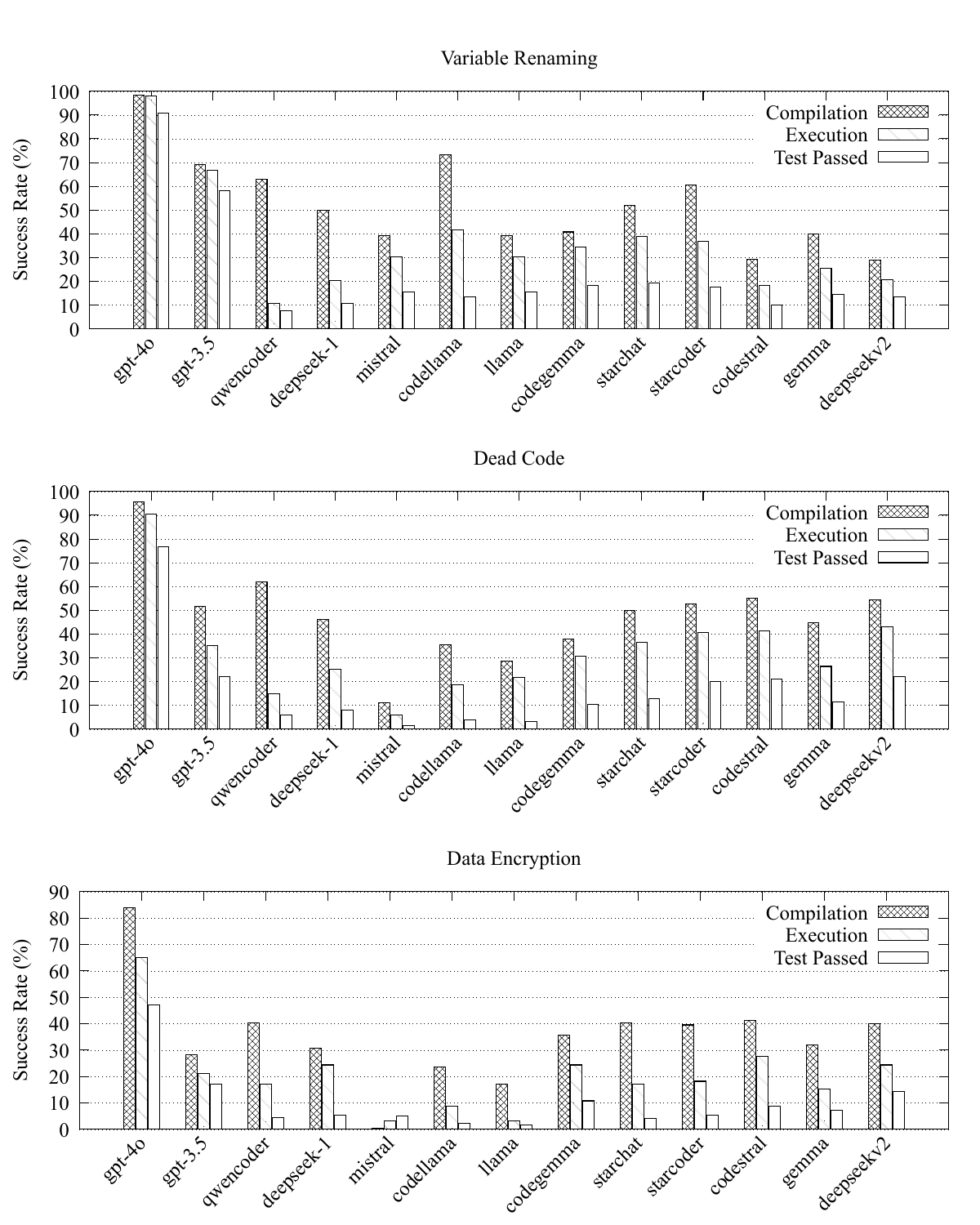} 
    % \vspace{-10pt}
    \caption{Evaluation of LLMs ability to deobfuscate code}
    \label{fig:Deobfuscation-all}
\end{figure}

\textbf{Results:}
Figure~\ref{fig:Deobfuscation-all} presents the results of our experiment. }
Our analysis reveals several important patterns:
\begin{description}[leftmargin=*]
\item \textit{Performance hierarchy across obfuscation types:} All models show a consistent pattern of decreasing performance according to obfuscation complexity: variable renaming presents the most approachable challenge, followed by dead code injection, with literal encryption proving most difficult. This hierarchy aligns with our findings on description generation tasks, suggesting consistent underlying limitations in LLM code processing.
\item \textit{Syntactic vs. semantic success:} We observe a substantial gap between compilation rates and functional correctness across all models and obfuscation types. For example, several models achieve 60-70\% compilation success on dead code obfuscation but only 40-50\% test-passing rates. This indicates that LLMs can often generate syntactically valid code that fails to preserve the original semantics.

\item \textit{Specialized vs. General models:} 
Interestingly, code-specia\-lized models generally outperform general-purpose ones on deobfuscation tasks (except for the GPT family), which contrasts with our findings for description generation tasks.
This discrepancy suggests that different aspects of code understanding are captured by these two task types, with deobfuscation potentially benefiting more from exposure to diverse code transformations during training.

\item \textit{Variable renaming resilience:} Several top-performing models achieve impressive results on variable renaming deobfuscation, with compilation rates above 60\% and test-passing rates reaching 90\% for the best models. This suggests that despite the strong impact of variable renaming on description tasks, LLMs can effectively reverse this transformation when explicitly prompted to do so.

\end{description}

\finding{6}{
Large Language Models for code struggle with code description tasks. However, they demonstrate superior performance when it comes to deobfuscation.
}

\subsubsection{Semantic Preservation in LLM Deobfuscation}
\textbf{Goal:} We evaluate whether LLM-deobfuscated code preserves the semantic intent of the original implementation. Understanding semantic preservation is crucial because functional equivalence alone does not guarantee that the code's underlying algorithmic approach and design patterns are maintained. 

\begin{figure}[htbp]
    \centering
    \includegraphics[width=.9\linewidth]{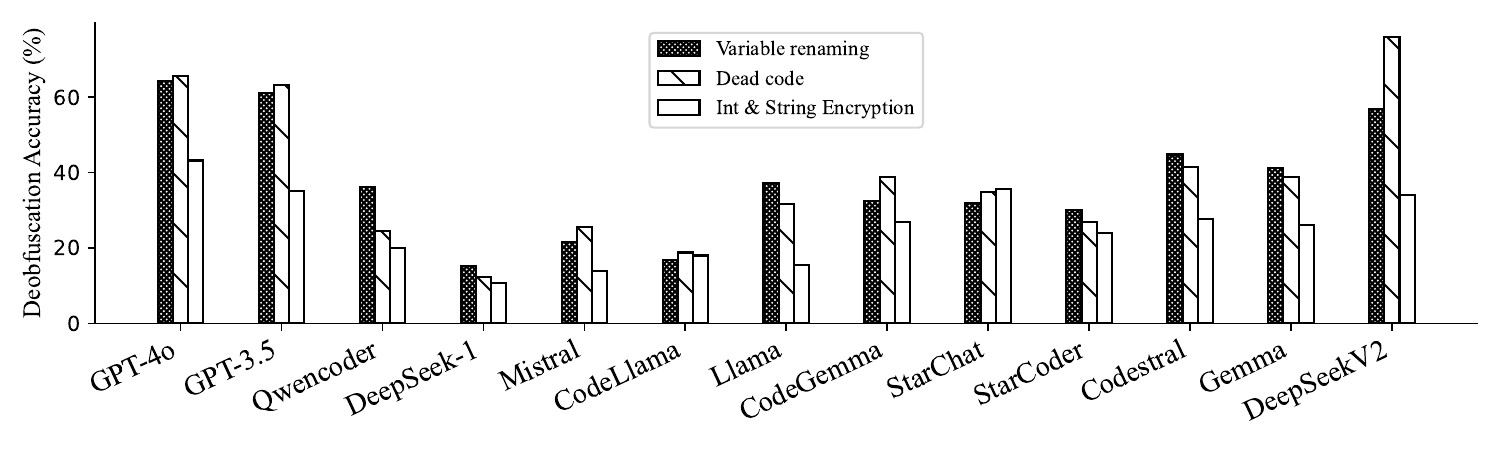} 
    \vspace{-10pt}
    \caption{Deobfuscation success rates across different obfuscation techniques}
    \label{fig:Deobfuscation-types}
\end{figure}

\noindent
\textbf{Experiment:} We analyzed code successfully deobfuscated by LLMs (code that compiled and passed functional tests) by generating descriptions of both the deobfuscated outputs and the original code using GPT-4o, then comparing their semantic similarity.

\noindent
\textbf{Results:} 
Figure~\ref{fig:Deobfuscation-types} shows deobfuscation performance across models and obfuscation types. 
We observe that models achieving high functional correctness (particularly GPT-4o, DeepSeekV2, and GPT-3.5) still vary significantly in their ability to restore semantic clarity.
Variable renaming proved most tractable (22.7\% average success), followed by dead code removal (16.3\%), with literal encryption presenting the greatest challenge (12.7\%). 
Manual analysis showed that some LLM-deobfuscated code works but does not fully restore original clarity, creating a misleading impression of deobfuscation success when evaluating solely through functional testing.

\finding{7}{
Even when LLMs produce deobfuscated code that passes functional tests, it often remains partially obfuscated. The code works but is not fully restored to its original clarity. LLMs are best at renaming variables, decent with numbers, but consistently struggle with decrypting strings. Additionally, specialized code LLMs generally perform better at deobfuscation tasks than general models, with GPT-4o standing as the notable exception that outperforms even specialized code models.
}

\section{Discussion}
\label{sec:discussion}
Throughout our discussion, we recognize that the ability to generate accurate code descriptions is only one indicator of code understanding. We now discuss the key implications and threats to validity.

\subsection{Key Findings and Implications}
Our study yields six significant findings regarding LLMs' capabilities in processing obfuscated code:

\noindent
\bcircle{1}
First, general-purpose LLMs (particularly GPT-4o and Deepseek-R1) outperformed code-specialized models in generating accurate descriptions of obfuscated code. 
This unexpected result challenges the assumption that domain-specific fine-tuning necessarily improves performance on code-related tasks. 
Instead, the breadth of training data and base model quality may be more important factors for robust code processing. 
Tasks related to obfuscated code analysis can therefore be considered as new tasks in the development of more related code-specialized LLMs.

\noindent
\bcircle{2}
Second, different obfuscation techniques revealed specific dependencies in how LLMs process code when generating descriptions. 
Variable renaming caused most significant performance degradation across all models, indicating a strong reliance on meaningful identifier names rather than structural understanding. 
Models showed surprising resilience to dead code injection, suggesting some ability to filter irrelevant information when producing descriptions. 
However, integer and string encryption significantly impaired description accuracy, revealing limitations in processing code when literal values are obscured or when the code is inflated with complex code (for dynamic decryption).

\noindent
\bcircle{3}
Third, our analysis of failure cases revealed that LLMs particularly struggle with describing concise yet structurally complex code. 
This combination of high information density and logical complexity represents a specific blind spot in current model architectures, which has implications for their application to highly optimized or production-grade code.

\noindent
\bcircle{4}
Fourth, description accuracy improvement when translating non-English comments highlights a practical constraint in multilingual code processing. 
This finding has important implications for international development teams and automated code analysis tools.
%\sn{Ok. I've noticed an inconsistency that I'm in the process of checking.}

\noindent
\bcircle{5}
Fifth, while code-specialized LLMs often fail to generate accurate descriptions, they generally outperform general-purpose models on deobfuscation tasks (with the exception of the GPT family). Description generation and deobfuscation appear therefore to engage different aspects of code understanding. Code-specialized models excel at structural transformations but struggle with abstracting and articulating code purpose in natural language. 

\subsection{Theoretical and Practical Implications}
The aforementioned findings suggest that current LLMs process code using a layered approach that includes surface-level pattern recognition, structural analysis, and semantic integration attempts. 
This processing model explains why no single obfuscation technique completely defeated all LLMs, while also explaining why all models showed some performance degradation in description tasks.

For software engineering practice, our results indicate that:
\begin{itemize}[leftmargin=*]
\item Meaningful variable names significantly improve LLMs' ability to process code, potentially even more than removing dead code or maintaining clear control flow.
\item LLMs may provide less reliable assistance when working with highly optimized, concise code bases compared to more verbosely written application code.
\item Language barriers in comments represent a meaningful constraint for current LLMs, suggesting the value of standardized documentation languages in international teams.
\end{itemize}

For security applications, the varied effectiveness of different obfuscation techniques provides insights for both software protection and reverse engineering. 
While LLMs show promise as assistive tools for deobfuscation tasks, their limitations necessitate human oversight, particularly when dealing with sophisticated obfuscation.

\subsection{Threats to Validity}
We identify several potential threats to the validity of our findings:

\noindent {\bf Internal Validity}

\noindent \textit{Proxy measurement limitations.} Description generation may not fully capture code comprehension. Models could generate plausible descriptions without understanding algorithms. We mitigate this by also examining deobfuscation performance.

\noindent \textit{Evaluation subjectivity.} Using LLMs to evaluate other LLMs' descriptions introduces potential circularity. We addressed this by using two different evaluation models with clear thresholds and manual verification of samples.

\noindent \textit{Prompt sensitivity.} LLM performance varies with prompt formulation. We used consistent, carefully designed prompts across all models and tasks, though different prompting strategies might yield different results.

\noindent {\bf External Validity}

\noindent \textit{Dataset characteristics.} Our dataset of algorithmic problems, in particular those stemming from coding competitions, may not represent the complexity and diversity of real-world software systems. However, these are extracted from widely-used benchmarks for assessing automated software engineering tools.

\noindent \textit{Language specificity.} 
% \tb{I guess it is all Java right?}\js{yes} 
Our Java-only focus limits generalizability to languages with different paradigms or structures. 
However, Java's widespread use, C-family syntax, and object-oriented paradigm make it representative of many mainstream languages. The obfuscation techniques we studied (variable renaming, dead code injection, and literal encryption) are largely language-agnostic and apply similar transformations across multiple languages. While language-specific nuances exist, the fundamental challenges of obfuscation likely transcend specific syntax.

\noindent \textit{Obfuscation techniques.} We used three common techniques rather than an exhaustive set. 
However, these techniques represent fundamentally different approaches to obfuscation (lexical, structural, and semantic) that target distinct aspects of code understanding. 
This coverage of the primary obfuscation categories provides insights likely to generalize to variations and combinations of these basic techniques.

\noindent 
{\bf Construct Validity}

\noindent \textit{Model selection.} Our selection of 13 models, while not exhaustive, strategically covers diverse architectures, parameter scales, and training approaches, including both widely used commercial models and leading open-source alternatives. 
The consistent patterns observed across this diverse set suggest our findings represent fundamental properties of current LLM technology rather than model-specific behaviors.

\subsection{Limitations and Future Directions}
Despite the raised threats, the consistent patterns observed in multiple models suggest that our findings are robust.
Future research should explore: 
\wcircle{1} enhancing LLMs' representation learning to reduce dependency on specific naming conventions; 
\wcircle{2} investigating whether adversarial training with obfuscated code improves robustness; 
\wcircle{3} developing hybrid approaches that combine LLM capabilities with traditional program analysis techniques; and
\wcircle{4} establishing more diverse proxies for code understanding beyond description generation.
The surprisingly strong performance of general-purpose models suggests that exposure to diverse text patterns during pre-training may be more valuable than narrow code specialization—an insight that could inform future approaches to building more robust code-processing LLMs.

\section{Conclusion}
\label{sec:conclusion}

This study systematically evaluated state-of-the-art LLMs' ability to process obfuscated code through description generation and deobfuscation tasks. Our findings reveal a critical dependency of LLMs on lexical elements rather than structural understanding. Contrary to expectations, general-purpose LLMs demonstrated greater resilience to obfuscation than code-specialized models. We also identified that all models struggle with concise yet structurally complex code—a blind spot with implications for real-world applications.

We show how code obfuscation serves as an effective challenge task that forces models beyond surface pattern recognition, exposing their reliance on lexical patterns rather than deeper semantic understanding. Developing models that can overcome the obfuscation frontier would represent a substantial advancement in representation learning for code, enabling semantic equivalence detection across syntactic variations and potentially revolutionizing automated program analysis

\bibliographystyle{ACM-Reference-Format}
\bibliography{references}

\end{document}